TITLE

# Proximity Induced Charge Density Wave in a Graphene/1T-TaS$_2$ Heterostructure

AUTHOR LIST

Nikhil Tilak[1+], Michael Altvater[1+], Sheng-Hsiung Hung[2+], Choong-Jae Won[3],

Guohong Li[1], Taha Kaleem[1], Sang-Wook Cheong[1], Chung-Hou Chung[4,5,6], Horng-Tay Jeng[2,5,7], and Eva Y. Andrei[1]

AFFILIATIONS

1 Department of Physics and Astronomy, Rutgers, the State University of New Jersey, 136 Frelinghuysen Rd, Piscataway, New Jersey 08854, USA

2 Department of Physics, National Tsing Hua University, 101 Kuang Fu Road, Hsinchu 30013, Taiwan

3 Laboratory for Pohang Emergent Materials and Max Plank POSTECH Center for Complex Phase Materials, Department of Physics, Pohang University of Science and Technology, Pohang 37673, Korea

4 Department of Electrophysics, National Yang Ming Chiao Tung University, 1001 University Rd., Hsinchu, Taiwan 300

5 Physics Division, National Center for Theoretical Sciences, Taipei 10617, Taiwan

6 Center for Theoretical and Computational Physics, National Yang Ming Chiao Tung University, Hsinchu 30010, Taiwan

7 Institute of Physics, Academia Sinica, Taipei 11529, Taiwan

*Corresponding authors E-mail: chung0523@nycu.edu.tw, jeng@phys.nthu.edu.tw, eandrei@physics.rutgers.edu

+Equal contributors

ABSTRACT

The proximity-effect, whereby materials in contact appropriate each other's electronic-properties, is widely used to induce correlated states, such as superconductivity or magnetism, at heterostructure interfaces. Thus far however, demonstrating the existence of proximity-induced charge-density-waves (PI-CDW) proved challenging. This is due to competing effects, such as screening or co-tunneling into the parent material, that obscured its presence. Here we report the observation of a PI-CDW in a graphene layer contacted by a 1T-TaS$_2$ substrate. Using scanning tunneling microscopy (STM) and spectroscopy (STS) together with theoretical-modeling, we show that the coexistence of a CDW with a Mott–gap in 1T-TaS$_2$ coupled with the Dirac-dispersion of electrons in graphene, makes it possible to unambiguously demonstrate the PI-CDW by ruling out alternative interpretations. Furthermore, we find that the PI-CDW is accompanied by a reduction of the Mott gap in 1T-TaS$_2$ and show that the mechanism underlying the PI-CDW is well-described by short-range exchange-interactions that are distinctly different from previously observed proximity effects.

## INTRODUCTION

The isolation and manipulation of atomically thin materials provides a ready-made two dimensional electron system[1] whose properties can be tuned by external knobs such as stress, substrate morphology or doping[2-8], to facilitate the emergence of correlated electron phases. Distinct from these external knobs, a powerful approach to manipulate electron correlations is by contact proximity effects. It is well known that proximitizing materials that host correlated electron phases with a normal metal, induces correlations in the metal[9]. This is a consequence of the quantum mechanical properties of electrons in solids; specifically, the non-local nature of electrons. As quantum particles do not have a well-defined position, electronic states cannot abruptly change from one type of ordering to another at the interface between two materials. Thus, correlated states persist into the normal metal where scattering events begin to destroy the coherence (and vice versa). In the case of 2D materials where scattering is reduced due to their atomically sharp interfaces, proximity-effects are particularly robust allowing proximitized states to persist over long distances. The discovery of graphene and other 2D materials, together with technology enabling the fabrication of 2D heterostructures has led to the observation of strong proximity effects at the atomic limit including proximity induced superconductivity, magnetism[10] and spin–orbit effects[11-18].

CDW states [19] [20] that are typically induced by Fermi surface instabilities driven by electron phonon coupling[21] or by correlation enhanced exchange effects[22,23], are quite robust. By contrast, PI-CDW in contacted 3D metals are fragile and easily destroyed by interface defect scattering. The use of 2D heterostructures fabricated in an inert atmosphere to avoid interface damage and contamination has made it possible to address this difficulty by substantially reducing interface scattering [24,25]. However, unambiguously identifying a PI-CDW in a 2D heterostructure is challenging due to competing mechanisms such as screening, doping or strain which can also produce charge-modulations. In addition, as is the case of local probes such STS, co-tunneling into the parent material can obscure the proximity induced CDW. Therefore, in order to demonstrate a PI-CDW one must: (i) demonstrate the presence of a CDW in the contacted material with the same period as that of the parent material (ii) rule out misidentification of electrostatic screening effects as a PI-CDW (iii) rule out misidentification of the parent CDW as the PI-CDW. Thus far studies of 2D heterostructures comprised of graphene and transition metal dichalcogenides (TMD) were unable to satisfy the three necessary criteria to unambiguously demonstrate the existence of a PI-CDW[24-27].

In this work, we present evidence of PI-CDW in a graphene layer contacted by a 1T-TaS$_2$ substrate hosting a commensurate CDW (CCDW). This discovery was made possible by the unique coexistence of the CDW and Mott gap in 1T-TaS$_2$ together with the Dirac spectrum of graphene. This exceptional combination, which was not present in any of the previously studied systems[24-27], enabled to unambiguously establish the presence of the PI-CDW by ruling out charge density modulations induced by trivial co-tunneling processes or by screening effects. By comparing STM and STS measurements with first-principle calculations as well as with mean-field theory results, we demonstrate that the charge density modulation of the CCDW in 1T-TaS$_2$ persists within the contacted graphene layer. Further, we identify the effect of charge transfer on the band structure of 1T-TaS$_2$ and reveal that the PI-CDW is described by a mechanism of short-range exchange interactions.

## RESULTS AND DISCUSSION

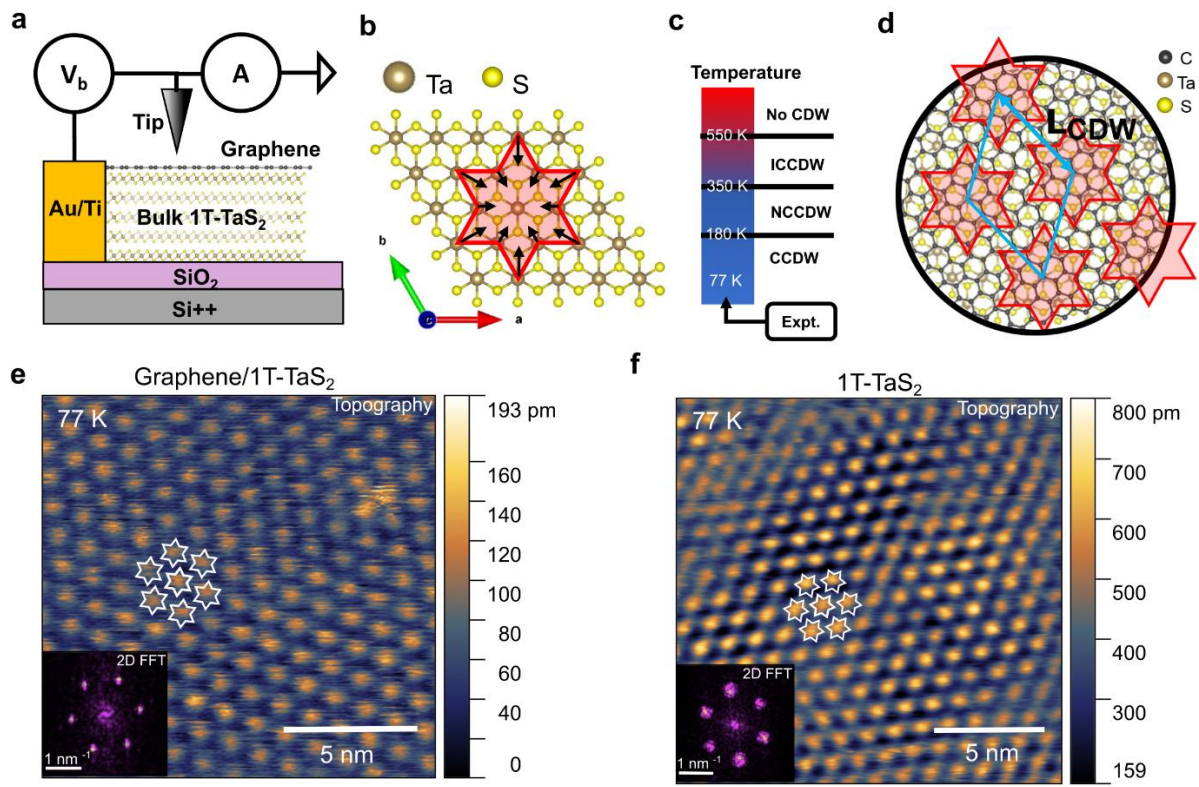

**Fig. 1 Scanning tunneling microscopy of graphene covered 1T-TaS$_2$.** (a) Schematic of the device and STM measurement setup. (b) Cartoon showing a single star-of-David (red) in 1T-TaS$_2$. The black arrows denote the displacement of Ta atoms towards a central Ta atom. (c) The transition temperatures between the incommensurate CDW (ICCDW), nearly commensurate CDW (NCCDW) and commensurate CDW (CCDW) phases of 1T-TaS$_2$. (d) Cartoon showing a monolayer graphene layer on top of 1T-TaS$_2$. Here, the angle between the graphene layer and 1T-TaS$_2$ is 13.9° to form a commensurate unit cell (cyan) for DFT calculations. (e) An STM topography image of graphene covered 1T-TaS$_2$ measured at 77 K using a tunneling set point of $I$=150 pA, $V_b$=250 mV. Each bright spot corresponds to a star of David with a measured CDW wavelength of 1.2 nm. This can also be seen in the 2D-FFT of the topography data (inset). (f) An STM topography image and its 2D FFT (inset) of a bulk 1T-TaS$_2$ sample measured at 77 K using a tunneling set point of $I$=150 pA, $V_b$=500 mV. The crystals in panels b and d were plotted using VESTA [49].

1T-TaS$_2$ is comprised of hexagonal layers of tantalum atoms coordinated octahedrally by sulfur atoms. Below 550 K a high temperature metallic phase transits into an incommensurate CDW phase, followed by a nearly commensurate CDW (NCCDW) below 350 K, and a CCDW below 180 K[24] (Fig. 1c). The CDW unit cell consists of a 13-atom cluster where 12 atoms displace from their high-temperature equilibrium positions toward the central, 13$^{th}$ Ta atom forming a $\sqrt{13} \times \sqrt{13}$ reconstructed supercell (Fig. 1b) also known as a David star (DS) structure. The CDW in 1T-TaS$_2$ is driven by strong electron-phonon coupling which critically suppresses an acoustic phonon mode along the $\Gamma$-$M$ direction[28,29] (Supplementary Fig. 4) leading to the static displacement of the lattice with the wavevector $\mathbf{Q_{CDW}}$, corresponding to the $\sqrt{13} \times \sqrt{13}$ CDW. This soft phonon mode consists primarily of longitudinal vibrations of Ta atoms with a minor contribution from transverse vibrations of S atoms relative to the phonon propagating direction $\mathbf{Q_{CDW}}$. The DS atomic arrangement involves the local lattice contraction around the center of the star, in which the bond lengths between Ta ions are shorter than those between Ta ions outside the DS.

We studied heterostructures of thin 1T-TaS$_2$ flakes (7-50 nm) covered by monolayer graphene (graphene/1T-TaS$_2$) using STM and STS (Fig. 1a). The samples were assembled inside a dry argon-filled glovebox (<0.1 ppm O$_2$, H$_2$O)[24,30] to avoid oxidation of the air sensitive surface of 1T-TaS$_2$ and subsequently cooled to 77 K inside a home-built STM[31,32]. At this temperature 1T-TaS$_2$ is in the CCDW regime. The STM topography (Fig. 1e) measured on a graphene/1T-TaS$_2$ sample using a positive bias set-point ($V_b$ = 250 mV) shows a triangular array of tall spots. The lattice spacing for this array is about 1.2 nm as measured directly from the topography image and confirmed with its fast Fourier transport (FFT) (Fig. 1e, inset). This spacing is equal to the expected CDW wavelength ($L_{CDW}$) in 1T-TaS$_2$ indicating that each tall spot corresponds to one DS. The topography of the graphene covered sample thus reflects the distinct arrangement of the DS clusters which are the hallmark of the CDW in 1T-TaS$_2$[33]. The topography is identical to that measured on a bulk 1T-TaS$_2$ sample without a graphene cover under similar tunneling conditions (Fig. 1f). This result also holds true for negative bias set points of $V_b < -300$ mV. In other words, for sufficiently large bias voltages, of either sign, there is no noticeable difference in the STM topography of 1T-TaS$_2$ with or without a graphene cover. This similarity might suggest that there is little or no interaction between the two materials. But spectroscopy and low bias-voltage topography scans show that this is not the case.

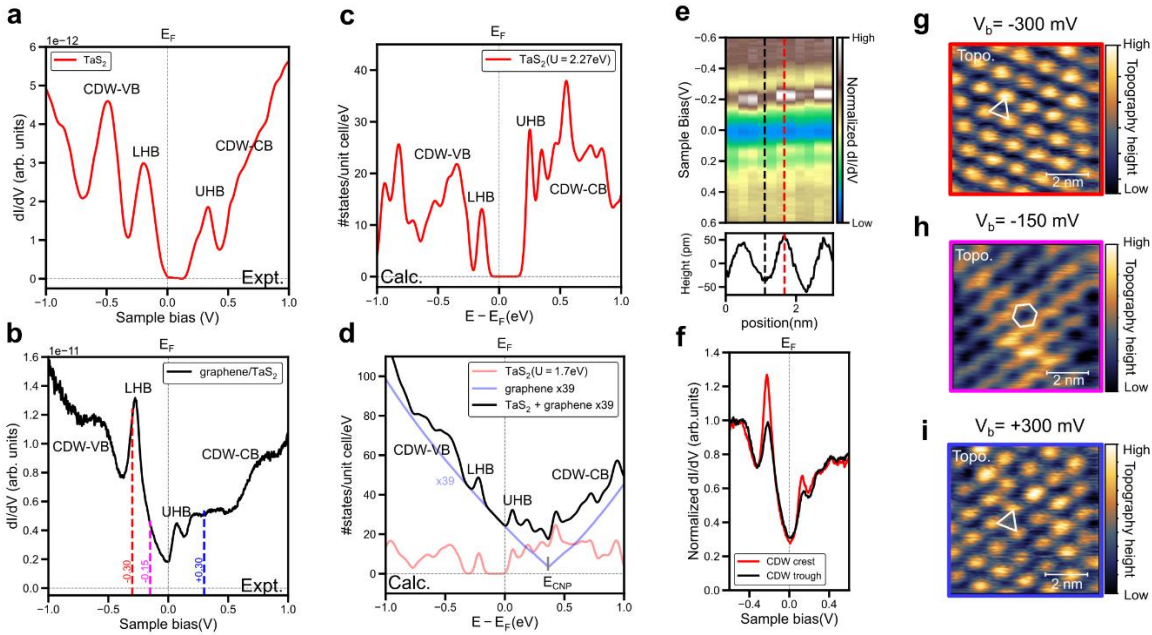

**Fig. 2 Effects of interaction between graphene and 1T-TaS$_2$.** (a) and (b) are the measured dI/dV curves on bulk 1T-TaS$_2$ and graphene/1T-TaS$_2$ respectively. The positions of the Hubbard band peaks and the broad features originating from the CDW distortion are labelled. The red, magenta and blue dashed lines in (b) indicate the bias voltages at which topography scans (g-i) were measured, respectively. (c) The calculated DOS for bulk 1T-TaS$_2$. (d) The calculated DOS for the heterostructure decomposed into the partial DOS of graphene (blue curve) and 1T-TaS$_2$ (red curve). The graphene DOS is multiplied by a factor (39) before adding the 1T-TaS$_2$ DOS (black curve) for comparison with experiment. This accounts for the smaller tip-graphene separation. See SI for how the factor was estimated. The linear dispersion around the Dirac point of graphene at ~0.3eV above $E_F$ indicates charge transfer from graphene to 1T-TaS2. (e) A map of *dI/dV (V$_b$)* vs position across 8 DSs in graphene/1T-TaS$_2$. The intensity of both Hubbard bands is highest at the center of the DSs. (f) Three representative spectra from the map in (e) demonstrate this intensity variation. (g-i) STM topography scans measured at bias voltages of -300 mV, -150 mV and +300 mV respectively indicated by dashed lines in (b). The tunneling resistance was kept constant at 1 GΩ for all scans. The small lateral mismatch between scans is a result of thermal drift of the sample. All scale bars are 2 nm.

The high temperature band structure of bulk 1T-TaS$_2$ consists of an isolated, half-filled spin-degenerate flat band at the Fermi level ($E_F$) (Supplementary Fig. 5). In the CCDW phase, the flat band splits into an occupied spin up lower Hubbard band (LHB) and empty spin down upper Hubbard band (UHB) with a Mott gap of ~0.5 eV in between (ab initio calculation Fig. 2c). The other bands corresponding to the CDW distortion appear at energies below the LHB (CDW-VB) and above the UHB (CDW-CB). These spectroscopic features are clearly observed in the experimental $dI/dV$ spectrum (Fig. 2a) measured on a bulk 1T-TaS$_2$ sample, in agreement with previous reports[20,23,34,35].

The spectral features of graphene/1T-TaS$_2$ (Fig. 2b), like those of bare 1T-TaS$_2$, show two peaks at ~ -270 meV and +70 meV identified as the LHB and UHB respectively. This identification is supported by the intensity variation of these features as a function of distance from the DS centers (Fig. 2e,f). The main differences between the 1T-TaS$_2$ and graphene/1T-TaS$_2$ spectra are the shift of the Fermi level and the linearly dispersing states within the Mott gap. The former can be understood in terms of the interfacial charge transfer caused by the work function difference between graphene ($\Phi_G$ ~ 4.6 eV [36]) and 1T-TaS$_2$ ($\Phi_{1T-TaS2}$ ~ 5.2 eV[37]). Consequently, the Fermi level moves from the LHB top to the UHB bottom, while the graphene Dirac point shifts to a higher energy about 0.3eV above $E_F$, indicating hole doping. The linearly dispersing Dirac cone states give rise to a linear band within the Mott gap of the 1T-TaS$_2$. While the DFT model reproduces the presence of graphene-attributed density of states at the Fermi level, we find that the experimental $dI/dV$ spectrum at higher energies is qualitatively different from the calculated DOS. The discrepancy arises from the exponential sensitivity of the signal on the distance between the STM tip and electronic states within the sample structure. Taking into account the decay of the wavefunction from the graphene surface to the 1T-TaS$_2$ substrate, we find that the contribution of the graphene layer to the calculated DOS should be multiplied by ~39 compared to that of 1T-TaS$_2$ layer (Supporting Information). The weighted sum (black) of the projected DOS of 1T-TaS$_2$ and that of graphene multiplied by 39 (blue) reproduces the asymmetric V-shaped DOS measured experimentally (Fig. 2d). While STM measurements of van der Waals heterostructures are common, de-convolving the different layer contributions to spectroscopy is not. Our procedure provides a computational method to accurately simulate STS measurements of 2D material heterostructures based on DFT models.

Another notable difference between the 1T-TaS$_2$ and the graphene/1T-TaS$_2$ spectra is that the separation between the Hubbard peaks in graphene/1T-TaS$_2$ (~338 mV) is smaller by about 31% compared to the bare 1T-TaS$_2$ (~491 mV). We attribute this to the reduction of the on-site Coulomb repulsion (*U*) in the graphene/1T-TaS$_2$ sample due to screening by the highly mobile carriers in graphene. These carriers suppress the localized picture of the narrow Hubbard d-bands in 1T-TaS$_2$ and thus reduce the on-site *U* value of Ta. Comparing the ab initio calculated band structures (supplementary Fig. 6) for different *U* values with the STS results, suggests that the Hubbard *U* is lowered from 2.27 eV to 1.70 eV. Using *U*=1.70 eV in our ab initio calculations (Fig. 2d) we find that the calculated DOS reproduces the main features of the observed STS. Our observation reveals important microscopic details about the effect of screening on correlated insulators which were not initially captured by the DFT model. Similar discrepancies between the measured and calculated DOS in heterostructures of graphene and sister TMDs (1T-NbSe$_2$ and 1T-TaSe$_2$[38-43]) may be resolved by taking into account the screening effects of the conducting substrates.

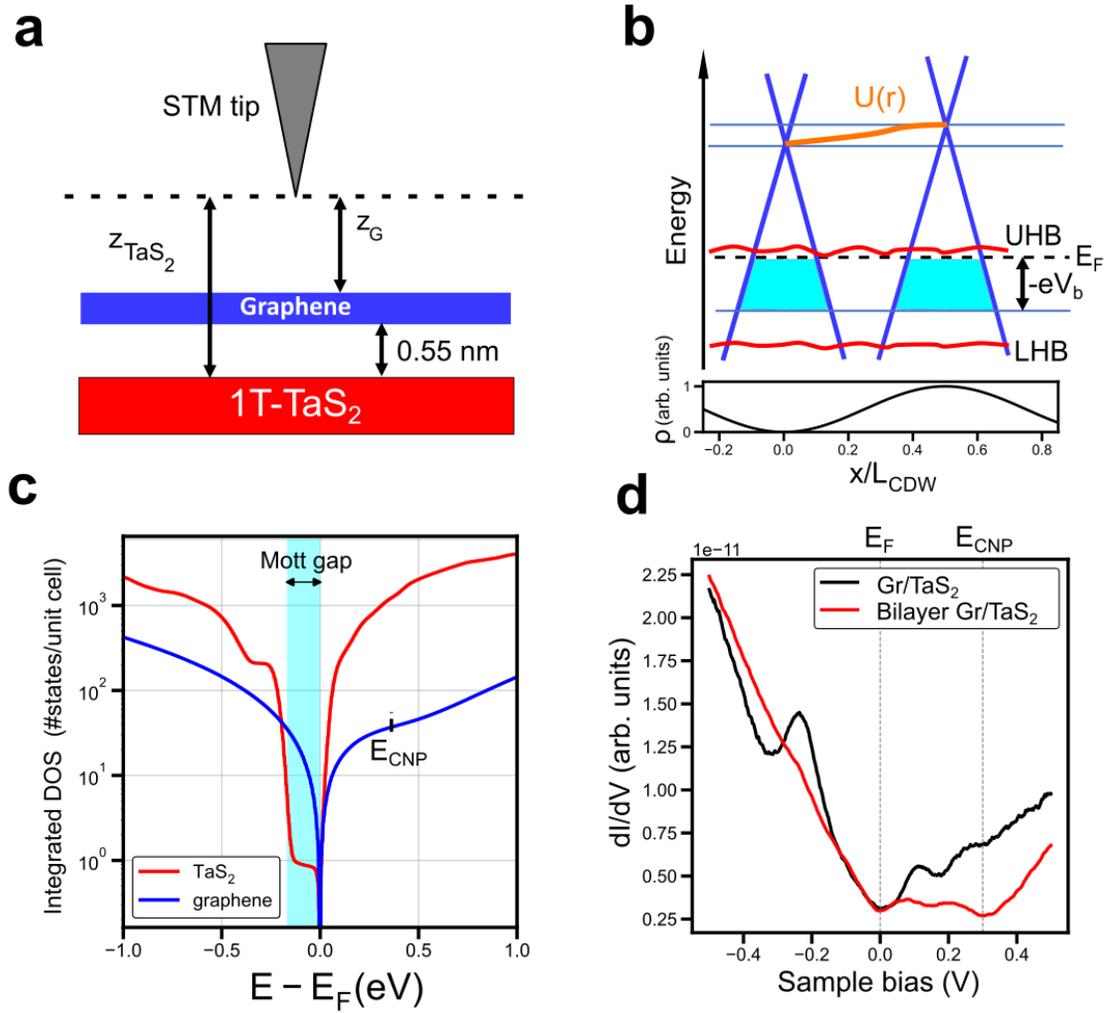

**Fig. 3 Modelling co-tunneling and electrostatic screening.** (a) A schematic cross section of the tunneling junction. The mean graphene-1T-TaS2 separation is ~ 0.55 nm. Co-tunneling into the graphene and 1T-TaS2 layers was estimated by measuring the tunneling current decay constant $z_0$. (b) The integrated DOS of the graphene (blue curve) and the 1T-TaS$_2$ (red curve) in the energy window $[E_F, E_F + eV_b]$ calculated using the DOS from DFT calculations. The integrated DOS for 1T-TaS$_2$ is about an order of magnitude larger than that of graphene for large $V_b$. For $V_b$ inside the Mott gap, the graphene DOS dominates. (c) A schematic which considers the effect of screening of the CDW in 1T-TaS$_2$ by the electrons in graphene. The bottom subplot shows the electron density ($\rho$) in the 1T-TaS$_2$ as a function of position. The resulting electric potential ($U(r)$) denoted by the orange line, tunes the local charge density of the graphene layer. This is denoted by the shift of the charge neutrality point of graphene's Dirac cones (blue) at two representative positions. The Fermi level is denoted by the dashed black line. The red squiggly lines denote the Hubbard bands of the 1T-TaS$_2$. The cyan region shows the states within the energy window $[E_F, E_F + eV_b]$ for the two positions. There are more states available for tunneling in the graphene above the center of the DSs compared to the sides. This should result in a triangular CDW like pattern in topography measured at these tunneling conditions. This is at odds with the observed hexagonal honeycomb pattern. (d) A *dI/dV* spectrum measured on a bilayer graphene/1T-TaS$_2$ area which shows that the CNP is clearly around 0.3 eV. The Hubbard bands are greatly suppressed compared to the monolayer graphene/1T-TaS$_2$ owing to the increased separation of the topmost graphene layer from the 1T-TaS$_2$, additional screening, and the lack of direct hybridization between the two materials.

In addition to the local height variation in the sample, STM topography also depends on the total number of states in the energy window $[E_F, E_F + eV_b]$. For CDW materials, the local DOS variation dominates the topography signal. Since the integrated DOS of 1T-TaS$_2$ is orders of magnitude larger than that of graphene for large $V_b$ (Fig. 3b), we estimate that there is substantial tunneling current into both the graphene and 1T-TaS$_2$, despite the larger z separation between the tip and the 1T-TaS$_2$ layer (Fig. 3a). Consequently, at high bias voltages the topography scans on graphene/1T-TaS$_2$ resemble those on 1T-TaS$_2$, as shown for $V_b$=-300 mV (Fig. 2g) and $V_b$=+300 mV (Fig. 2i). Note that these two energy windows include the LHB and UHB of the 1T-TaS$_2$.

Surprisingly, the topography pattern measured at bias voltages within the Mott gap of 1T-TaS$_2$ is hexagonal instead of the triangular pattern observed outside the Mott gap, but with the same period (Fig. 2h). We note that for bias voltages within the Mott gap of 1T-TaS$_2$ the STM topography includes only states belonging to graphene (Fig. 3b) because tunneling into the 1T-TaS$_2$ layer is disallowed. Therefore, since a hexagonal lattice is the dual of a triangular lattice, we conclude that the electronic density redistribution in graphene due to its proximity to 1T-TaS$_2$ forms a CDW pattern which is out-of-phase with the CDW in the underlying 1T-TaS$_2$. This result excludes an interpretation of the data in terms of co-tunneling into the 1T-TaS$_2$ substrate. The hexagonal pattern was observed for multiple energy windows inside the 1T-TaS$_2$ Mott gap but never outside it (Supplementary Fig. 10).

Note that this out-of-phase CDW cannot be explained by a simple screening picture (Fig. 3c). To wit: if one models the CDW in 1T-TaS$_2$ as a periodic modulation of electron density ($\rho(r)$) which peaks at the center of the DSs (Fig. 3c, bottom), then the electrostatic potential ($U(r)$) near the surface of 1T-TaS$_2$ will also peak near the center of the DSs (orange line in Fig. 3c). A graphene layer placed on top of 1T-TaS$_2$ will experience this spatially varying potential leading to spatial doping variation, i.e., shifting of the local CNP. This CNP variation is in addition to the average shift of the CNP (~ 0.3 eV) due to work function difference between the two materials as discussed above. If we consider an energy window in the Mott gap of the 1T-TaS$_2$, then there are more graphene states (cyan regions in Fig. 3b) near the center of the DS compared to the region between neighboring DSs. This should result in a triangular CDW pattern in STM topography, which is in-phase with the underlying 1T-TaS$_2$ CDW. This is contrary to our observation of a hexagonal pattern, thus ruling out this simple picture. These results provide

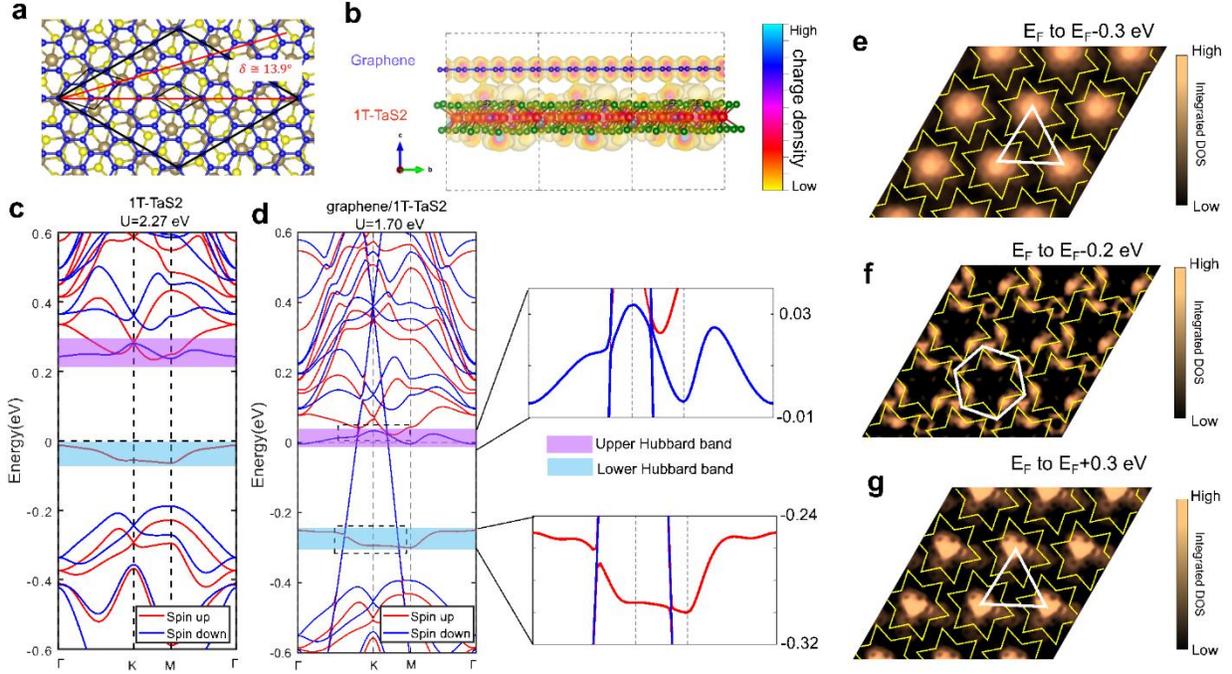

**Fig. 4 Ab initio band structure calculations.** (a) Top view of graphene/1T-TaS$_2$. Blue, brown, and yellow spheres indicate C, Ta, and S atoms, respectively. Black, blue, and brown rhombuses show the 5x5 graphene/$\sqrt{13} \times \sqrt{13}$ 1T-TaS$_2$ supercell, graphene 1x1 unit cell, and 1T-TaS$_2$ 1x1 unit cell, respectively. The calculations were carried out for a ~13.9° twist angle between graphene and 1T-TaS$_2$. (b) Side view of graphene/1T-TaS$_2$ overlaid with the charge density map (bubble-shaped color contour indicating the electron density) corresponding to the states at the two crossing points of the Dirac cone and Lower Hubbard band. The slight overlap between graphene and 1T-TaS$_2$ electron clouds gives rise to the interlayer coupling and proximity effect. Here, blue, red, and green spheres represent C, Ta, and S atoms, respectively. (c) GGA+U spin resolved band structure of $\sqrt{13} \times \sqrt{13}$ CDW reconstructed 1T-TaS$_2$ with $U$=2.27eV (d) GGA+U spin resolved band structure of graphene/1T-TaS$_2$ using the phenomenological value $U$ = 1.70 eV. Red and blue curves are the spin up and down bands, respectively. Owing to the charge transfer from graphene to 1T-TaS$_2$, the Fermi level (zero energy) moves from the lower Hubbard band to the upper Hubbard band, and the graphene-associated Dirac point is shifted to ~ 0.3 eV above the Fermi level indicating hole doping. Insets on the right are zoomed in views of the bands near the upper and lower Hubbard bands. (e-g) The integrated DOS of graphene/1T-TaS$_2$ in the energy range from E$_F$ to -300 mV, -200 mV and +300 mV, respectively, relative to that of freestanding graphene in corresponding energy ranges. DSs are overlaid for clarity. For the energy above the UHB (e) and below the LHB (g), the integrated DOS forms a triangular lattice. While within the Mott gap (f), the in-gap DOS forms a hexagonal lattice, in good agreement with our experimental observations in Fig. 2(g-i).

strong evidence for the existence of a PI-CDW in graphene.

We also studied an area of the sample covered with bilayer graphene. In this area the topography showed almost no CDW pattern. *dI/dV* spectra (Fig. 3d) showed a bilayer graphene-like V shape with a pronounced dip near 0.3 V, indicating the CNP. Very weak features corresponding to the position of the Hubbard bands were also seen. These observations indicate that the PI-CDW relies on very short-range coupling between the graphene and 1T-TaS$_2$.

We performed first-principle electronic structure calculations using the projector augmented wave (PAW) approach within the framework of density functional theory (DFT) as implemented in the Vienna ab initio Simulation Package (VASP)[44,45]. The exchange correlation is described in the Perdew-Burke-Ernzerhof (PBE) form of generalized gradient approximation (GGA)[46,47]. To take the strong correlation of Ta d-electrons into consideration, we perform GGA plus on-site $U$ (GGA+$U$) calculations with $U$=2.27eV for monolayer 1T-TaS$_2$ in accordance with previous DFT calculations[33,48]. The lattice structure is theoretically optimized with the atomic forces converged within 0.01 eV/Ang.

Next, we calculated GGA+U band structure of a monolayer 1T-TaS$_2$ covered with graphene using the reduced value $U$=1.7 eV (Fig. 4d) to reproduce the observed spacing between the Hubbard bands. To enable this calculation, we considered a twist angle of 13.9 degrees between the graphene and 1T-TaS$_2$ layers. This forms a commensurate superstructure (Fig. 4a) because $5 \times a_G \approx L_{CDW}$, where $a_G$=0.246 nm is the lattice constant of graphene (Supplementary Fig. 8). We note that both the 1T-TaS$_2$ Hubbard bands and the graphene Dirac cone are preserved (Fig. 4d) in the band structure. At the intersections between the Dirac cone with the Hubbard bands four small gaps ~ 10meV emerge (Fig. 4d insets). These gaps, which originate from the weak interlayer couplings, indicate finite hybridization between the Dirac and Hubbard states, giving rise to the proximity effect between graphene and 1T-TaS$_2$ as illustrated in Fig. 4b. In addition, we found a gap associated with the CDW distortion (SI Fig. 9), which was too small (~ 1.7meV) to detect experimentally.

Since different carbon atoms in the heterostructure unit-cell have a different registry with respect to the underlying CDW, it is natural to expect that the hybridization between graphene $p_z$-bands and 1T-TaS$_2$ $d$-bands varies as a function of position within the supercell. This can be seen in the calculated charge density map (Fig. 4b). To show the charge modulation of graphene in the proximity CDW state (Fig. 4e-g) for comparison with the experimental topography maps (Fig. 2h-g), we calculate the integrated DOS in the energy windows *[E$_F$, E$_F$ + eV$_b$]* for *V$_b$* =-0.3 V, -0.2 V and +0.3 V by subtracting the counterparts of freestanding graphene in the corresponding energy ranges, i.e., in the energy windows [-0.3 eV, -0.6 eV], [-0.3eV, -0.5 eV] and [-0.3 eV, 0.0 eV] with 0.0 eV being the Dirac point (*E$_F$*) of freestanding graphene. Consistent with the experimental results, we find triangular, hexagonal, and triangular CDW patterns below, within, and above the Mott gap region, respectively, providing further support for the out-of-phase PI-CDW in the graphene layer.

The PI-CDW is an effect driven by periodically varying band hybridization. In addition to the DFT calculations, we further modeled the graphene/1T-TaS$_2$ system using a simplified mean field Hamiltonian (detailed in the SI) based on the existing CDW order in 1T-TaS$_2$ with a weak charge transfer between the graphene and 1T-TaS$_2$ layers. The charge transfer is described by a short-ranged exchange term produced by second-order perturbation of the weak interlayer hopping. Using a mean-field decoupling of this exchange term, a PI-CDW order (much weaker than that in 1T-TaS$_2$), is generated in the graphene layer. We find that the CDW in graphene is out of phase with that in the 1T-TaS$_2$ layer, consistent with both the DFT and the experimental results. In addition, the mean-field theory predicts that the amplitude of the CDW in graphene is much weaker than that in 1T-TaS$_2$, again consistent with the DFT and the STM results. These findings provide strong support for the CDW proximity effect in graphene. We emphasize that the above mechanism based on charge transfer is distinct from previously realized proximity effects, including superconducting, magnetic, and spin–orbit proximity effects.

In summary, using STM/STS together with DFT calculations and mean field Hamiltonian to study graphene/1T-TaS$_2$, we demonstrated the existence of a PI-CDW in graphene and elucidated details of the coupling between the two systems. This was made possible by the unique properties of these two materials, the coexistence of the CDW and Mott gap in 1T-TaS$_2$ together with the Dirac spectrum of electrons in graphene, which enabled us to distinguish the PI-CDW in graphene from the contribution of the host CDW, as well as to rule out a screening-induced charge modulation. We propose a model based on the short-range exchange interaction between carriers in graphene and in 1T-TaS$_2$ which captures the main features of the CDW proximity effect. Concomitant with the PI-CDW in graphene, we observe a reduction in the Mott gap in 1T-TaS$_2$, indicating the presence of proximity-induced mid-gap carriers which screen the Mott–Hubbard interaction. The graphene/1T-TaS$_2$ system provides a sensitive and discriminating probe of contact-induced contributions such as charge transfer, screening, and exchange interactions at the atomic level. Future studies using techniques such as STM/STS and spin resolved STM to focus on these effects in heterostructures comprised of TMDs and metallic layers, could provide new insights into correlated insulator physics and quantum interactions at vdW interfaces, crucial for identifying emergent phenomena, and for understanding quantum transport through 2D materials embedded in 3D structures.

## METHODS

### Sample fabrication and STM measurements.

Samples were fabricated by mechanical exfoliation of graphene and separately 1T-TaS$_2$ flakes inside an argon-filled glovebox. 1T-TaS$_2$ flakes were exfoliated from bulk 1T-TaS$_2$ crystals (samples were purchased from 2D Semiconductors or grown by iodine chemical vapor transport) and transferred onto a passivated SiO$_2$-capped degenerately doped Si wafer. The graphene and 1T-TaS$_2$ flakes were aligned vertically and brought into close contact with micromanipulators under an optical microscope and then heated to promote adhesion. Standard electron beam lithography and electrode deposition (4-5nm Ti/40-50nm Au) were used to make electrical contact to the sample. After removing the PMMA mask the resulting heterostructure was annealed (180−220°C) in hydrogen/argon (10% : 90%) to remove polymer residues followed by AFM tip sweeping. STM and STS were performed using a homebuilt STM at 77 K in high-vacuum < $10^{-5}$ Torr. To locate the micron size samples we employed a technique using the STM tip (mechanically cut Pt/Ir ) as a capacitive antenna[32]. Two such samples were measured by STM using several mechanically cut Pt/Ir wire tips. Additionally, we also studied a 1T-$TaS_2$ flake transferred to a pre-patterned gold electrode using PDMS polymer. This sample was transferred to the STM system with < 15 mins of exposure to the atmosphere.

## DATA AVAILABILITY

The experimental data generated in this study have been deposited in the Zenodo database under accession code ZZ https://doi.org/10.5281/zenodo.12745631.

## ACKNOWLEDGEMENTS

NT and EYA acknowledge support from the Department of Energy grant DOE-FG02-99ER45742 and the Gordon and Betty Moore Foundation EPiQS initiative grant GBMF9453; MAA was supported by the National Science Foundation grant EFRI 1433307; GL was supported by Rutgers University; CJW was supported by The National Research Foundation of Korea (NRF), Ministry of Science and ICT(No. 2022M3H4A1A04074153); SWC was supported by the center for Quantum Materials Synthesis (cQMS), funded by the Gordon and Betty Moore Foundation's EPiQS initiative grant GBMF10104, and by Rutgers University, C-H C was supported by MOST (Grant NO.: 107-2112-M-009-010-MY3, 110-2112-M-A49-018-MY3),  NSTC (Grant NO. 112-2124-M-A49-003), and NCTS of Taiwan, R.O.C., JHT acknowledges support from the Ministry of Science and Technology, Taiwan under grant: MOST 109-2112-M-007 -034 -MY3, and from NCHC, CINC-NTU, AS-iMATE-109-13, and CQT-NTHU-MOE, Taiwan.

## AUTHOR CONTRIBUTIONS

N.T. and M.A. built the STM, performed the experimental studies, analysis, and wrote the paper; S-H.H and H-T.J. performed the first principles calculations; C-J.W grew one of the TaS2 samples; G.L. built the STM and helped with data collection; T.K. helped with sample fabrication; S-W.C. helped obtain the TaS2 sample; C-H.C. carried out the mean field calculations; E.Y.A. directed the project, analyzed data, and wrote the paper.

## COMPETING INTERESTS
The authors declare no competing interests.

# Supplementary information for "Proximity induced charge density wave in graphene/1T-TaS$_2$."

Nikhil Tilak[1+], Michael A. Altvater[1+], Sheng-Hsiung Hung[2+], Guohong Li[1], Choong-Jae Won[3], Taha Kaleem[1], Sang-Wook Cheong[1], Chung-Hou Chung[4,5*], Horng-Tay Jeng[2,6,5*] and Eva Y. Andrei[1*]

[1] Department of Physics and Astronomy, Rutgers, the State University of New Jersey, 136 Frelinghuysen Rd, Piscataway, New Jersey 08854, USA

[2] Department of Physics, National Tsing Hua University, 101 Kuang Fu Road, Hsinchu 30013, Taiwan

[3] Laboratory for Pohang Emergent Materials and Max Plank POSTECH Center for Complex Phase Materials, Department of Physics, Pohang University of Science and Technology, Pohang 37673, Korea

[4] National Chiao Tung University, 1001 Daxue Road, Hsinchu 30010, Taiwan

[5] Physics Division, National Center for Theoretical Sciences, Taipei 10617, Taiwan

[6] Institute of Physics, Academia Sinica, Taipei 11529, Taiwan

*Corresponding authors E-mail: chung0523@nycu.edu.tw, jeng@phys.nthu.edu.tw, eandrei@physics.rutgers.edu

+Equal contributors

## CONTENTS

# SAMPLE FABRICATION AND AFM SCANS

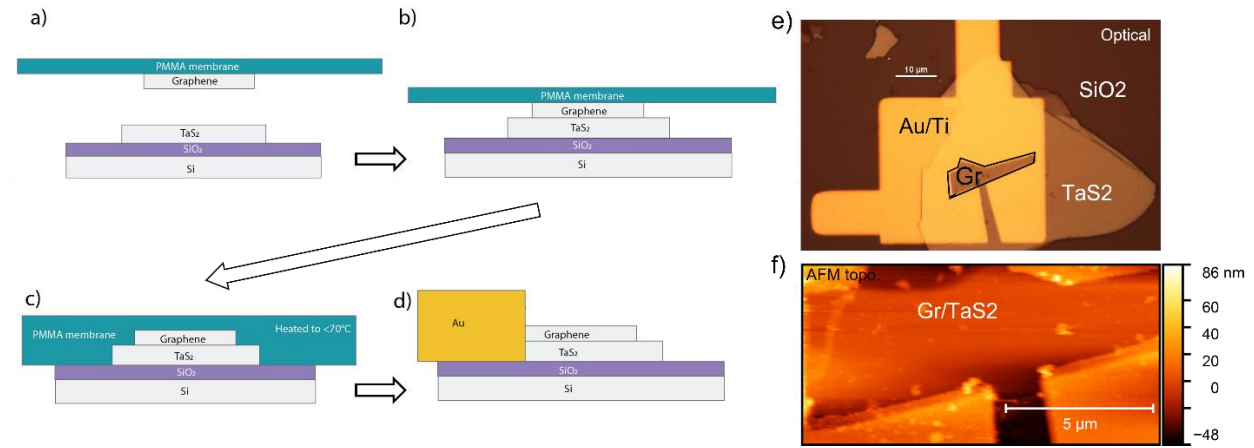

**Supplementary Figure 1: Graphene/ TaS$_2$ sample fabrication (a-d)** Schematic showing Graphene/ TaS$_2$ sample fabrication using a PMMA based wet transfer method. Steps a-c done in argon-filled glovebox. a) TaS$_2$ exfoliated onto SiO$_2$ chip and Graphene exfoliated onto PMMA membrane. Brought into alignment with micromanipulators in optical microscope setup. b) TaS2 and graphene brought into contact with micromanipulators. c) PMMA membrane melted onto SiO$_2$ chip. d) Gold electrode added using electron beam lithography. (e) Optical microscope image of completed sample. (f) AFM topography scan of sample collected just before loading into the STM.

Samples were prepared in an argon filled glovebox using an optical microscope setup and remote-controlled micromanipulators in a process following previous work[1]. First, graphite was exfoliated onto a polymethyl methacrylate (PMMA) membrane using tape, and an appropriate graphene flake was found using an optical microscope. Likewise, a bulk flake of TaS$_2$ was exfoliated onto a Silicon Dioxide (SiO$_2$) chip and an appropriate flake was found (Appropriate flakes are >20 µm along longest axis). This was done inside an Argon filled glovebox to prevent the oxidation of the TaS$_2$ flake. Using an optical microscope and remote-controlled micromanipulators, graphene and TaS$_2$ flakes were aligned (**Supplementary Figure** 1a) and brought into direct contact (Supplementary Figure 1b). Using a stage heater, the PMMA membrane was melted (T < 70 C) onto the SiO$_2$ chip (**Supplementary Figure** 1c), ensuring good contact between the graphene and TaS$_2$ flakes.

Since the Graphene cover protects TaS$_2$ from oxidation, the SiO$_2$ chip with the Graphene/TaS$_2$/PMMA combination was moved out of the glovebox to form a sample contact electrode(50 nm Au/4 nm Ti) using standard electron beam lithography, deposition, and liftoff techniques [1][2](**Supplementary Figure** 1d).

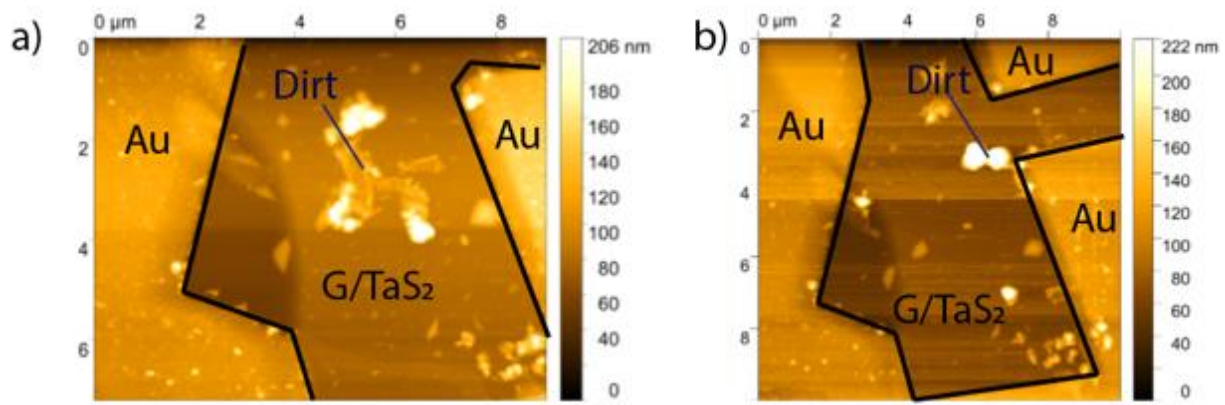

**Supplementary Figure 2: Residue removal via AFM sweeping.** Resist residues(dirt) left on the sample surface post annealing were swept to the edges of the sample using repeated scanning with an AFM tip in contact mode. (a) shows the sample before the sweeping steps and (b) shows the same area after a few contact mode scans.

The final Graphene/TaS$_2$ (G/TaS$_2$) heterostructure goes through a 2-stage cleaning process. First, we follow a standard procedure[1] of annealing the heterostructure in forming gas (10% Hydrogen in Argon) for over 12 hours at T=230 C. This removes the bulk of the resist residues. Next, we use an AFM contact mode topography sweeping technique on the Graphene surface [3] to move any dirt and residues to the edges of the graphene surface (**Supplementary Figure** 2).

Finally, the heterostructure is loaded into a homebuilt STM/STS system [4] [5] and a simple capacitive technique[6] is used to identify and scan the graphene surface.

A bare TaS$_2$ device was also made inside the glovebox using a peel-off process (**Supplementary Figure** 3a-b). Prepatterned electrodes were made on a Silicon wafer using a standard lithography technique. TaS$_2$ was exfoliated on a PDMS piece inside the glovebox. A suitable flake was identified under the optical microscope and aligned with an edge of the prepatterned electrode. Upon contact, the TaS$_2$ flake sticks to the rough SiO$_2$ substrate and detaches from the PDMS. Electrical contact is established via the overlap between the TaS$_2$ flake, and the electrode as seen in the optical microscope image (**Supplementary Figure** 3c). The sample is then loaded onto a sample flag inside the glovebox. It is then taken out of the glovebox, wire bonded, transferred to the STM and pumped down to high vacuum. The total exposure time is <15 min. The sample shows a CCDW pattern in STM topography at 77 K (**Supplementary Figure** 3d). The surface contamination is due to partial oxidation of the surface after a brief exposure to air during the sample-loading process.

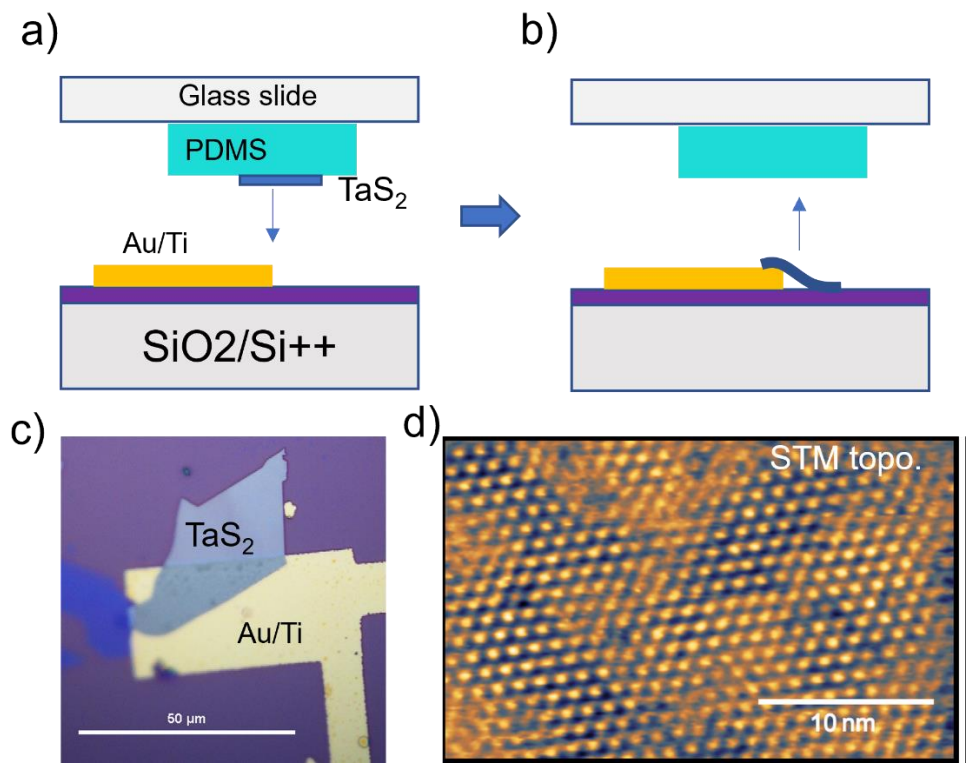

**Supplementary Figure 3: Bare TaS$_2$ sample (a)** TaS$_2$ exfoliated on PDMS and aligned with the edge of a pre-patterned Au/Ti electrode on a Silicon chip. **(b)** The TaS$_2$ is brought into contact with the electrode and retracted slowly, leaving the TaS$_2$ flake on the Silicon wafer. **(c)** Optical microscope image of the sample. **(d)** Example topography scan measured at Vb=500 mV and I=150 pA at 77K showing a CCDW pattern. The surface contamination is due to partial oxidation of the surface after a brief exposure to air during the sample-loading process.

# BAND STRUCTURE OF 1T-TAS$_2$

## Kohn Anomalies and CDW Supercell

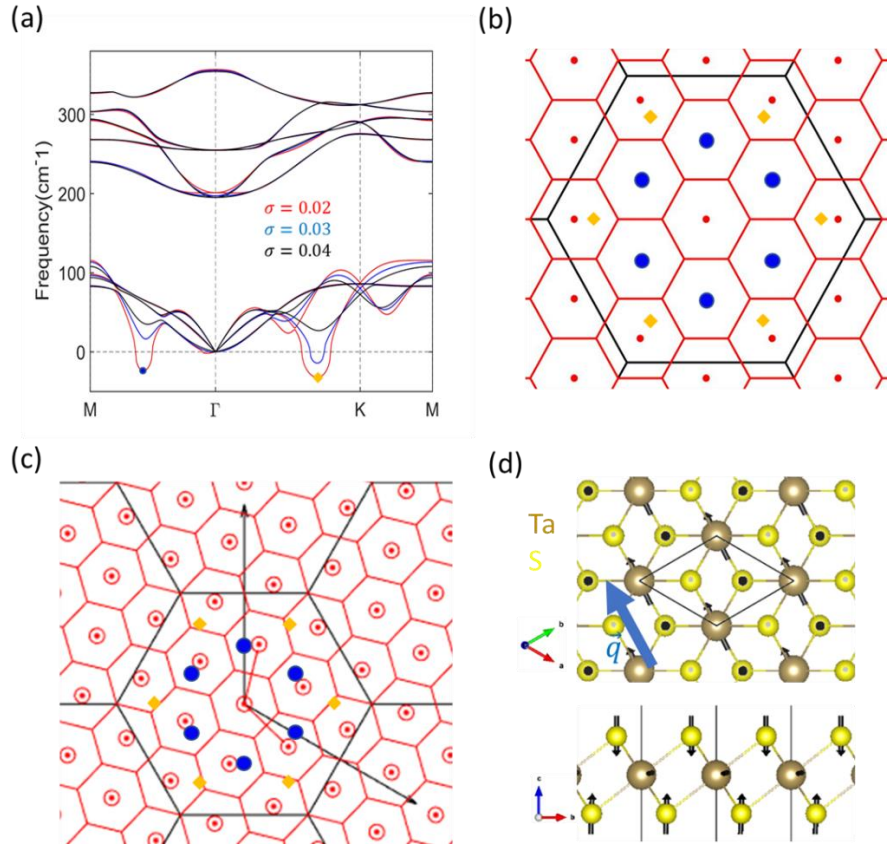

**Supplementary Figure 4: Kohn anomaly and CDW supercell (a)** Phonon dispersion of undistorted 1T-TaS$_2$. The Kohn anomaly wavevectors are marked by blue circle and yellow rhombus. Different broadening values σ indicate the temperature effect. **(b)** Kohn anomaly wavevectors and Brillouin zone (BZ) of $1 \times 1$ (black) and $\sqrt{13} \times \sqrt{13}$ (red) structures. The blue Kohn anomaly wavevector q along ΓM locates at the zone centers of the $\sqrt{13} \times \sqrt{13}$ CDW BZs, whereas the yellow Kohn anomaly wavevector along ΓK does not fit the $\sqrt{13} \times \sqrt{13}$ CDW BZs. **(c)** The $\sqrt{13} \times \sqrt{13}$ CDW BZs are rotated to form the commensurate BZs. **(d)** Ta atoms vibrate parallel to the Kohn anomaly wavevector q, while S atoms vibrate perpendicularly to q.

    The calculated phonon dispersion for the unreconstructed 1T-TaS$_2$ in Supplementary Figure 3a shows two Kohn anomalies, phonon modes whose frequencies drop to (or below) zero, indicating a lattice instability. These Kohn anomalies can be stabilized without negative frequency by using higher broadening (σ) values. This corresponds to the removed CDW phase at higher temperatures observed in experiments. The blue Kohn anomaly instability at wavevector **q** coincides with the wavevector **Q**$_{CDW}$ of the $\sqrt{13} \times \sqrt{13}$ CDW reconstruction, suggesting electron-phonon coupling as a possible origin for the CDW. Supplementary Figure 3b demonstrates that the blue Kohn anomaly wavevector **q** along ΓM locates at the zone centers of the $\sqrt{13} \times \sqrt{13}$ CDW BZs. The good match in the periodicity indicates the tendency from 1x1 unit cell toward the $\sqrt{13} \times \sqrt{13}$ CDW lattice. Although the q vector of 1x1 unit cell matches the period of $\sqrt{13} \times \sqrt{13}$ BZ, it is not possible for the 1x1 BZ to match the non-integer $\sqrt{13} \times \sqrt{13}$ BZ. This suggests a ~13.9° rotation of the $\sqrt{13} \times \sqrt{13}$ BZ as shown in Supplementary Figure 3c. The rotated $\sqrt{13} \times \sqrt{13}$ supercell thus commensurate with the 1x1 unit cell as observed and discussed in this

work experimentally and theoretically. On the other hand, the yellow Kohn anomaly does not fit the BZ of possible minimum supercells and thus is not observed. The blue soft phonon mode at the Kohn anomaly wavevector **q** contains mainly longitudinal vibration of Ta atoms with minor transverse vibration of S atoms relative to the phonon propagating direction **q** as depicted in Supplementary Figure 3d.

## CDW Induced Isolated Flat Band

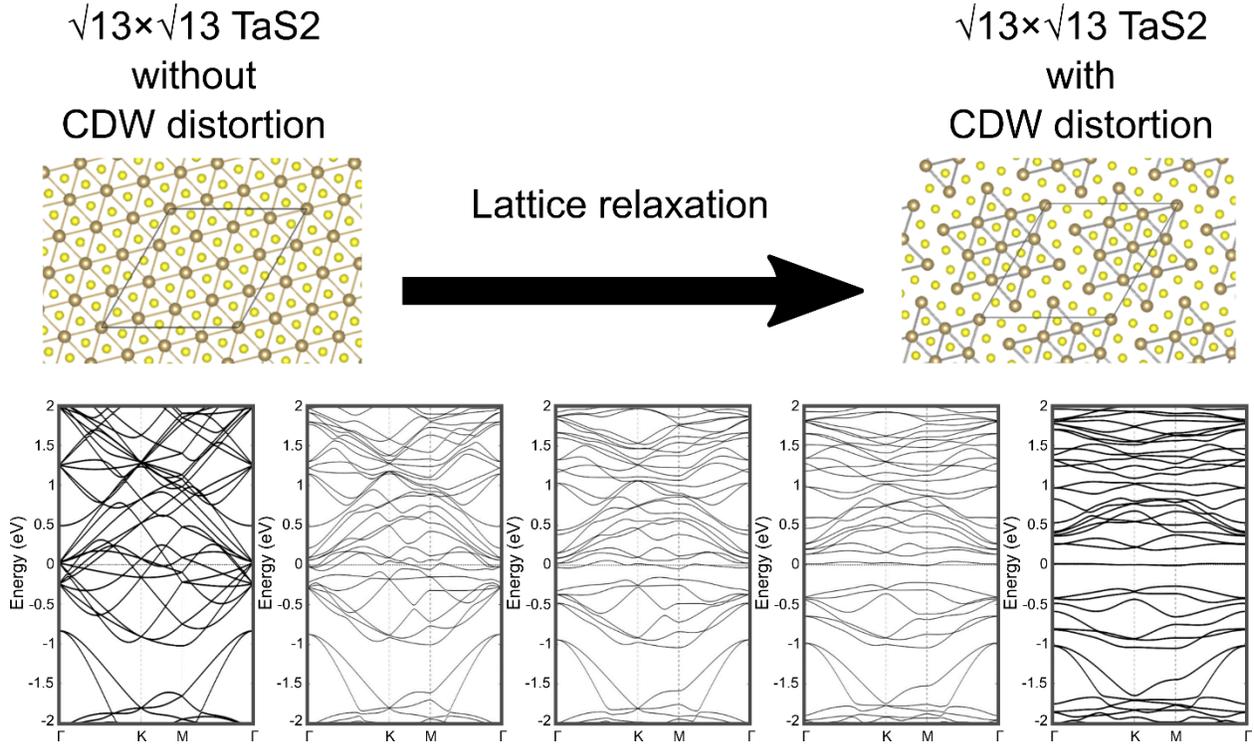

**Supplementary Figure 5: Evolution of band structure due to CDW distortion.** Evolution of the GGA band structure (bottom) as the lattice relaxes (from left to right) from the undistorted lattice structure (shown top, left) to the $\sqrt{13} \times \sqrt{13}$ distortion (top, right).

The undistorted 13 Ta atoms within the $\sqrt{13} \times \sqrt{13}$ supercell (top-left panel), the same size as the CDW unit cell (top-right panel), each contribute one conduction electron to the undistorted band structure (bottom-left panel). As the lattice is allowed to relax, 6 of these Ta-associated bands are pushed below the Fermi level while 6 are pushed to higher energies, leaving a single, flat, isolated band at the Fermi level (bottom-right panel). This band is associated with an electron state localized at the center of the Stars of David in the CDW reconstructed structure (top-right panel). As a flat, isolated band, the electronic kinetic energy (and kinetic energy spread) is low, allowing interparticle interactions to dominate and correlated electronic phases can emerge.

# Hubbard U Enhanced Gap

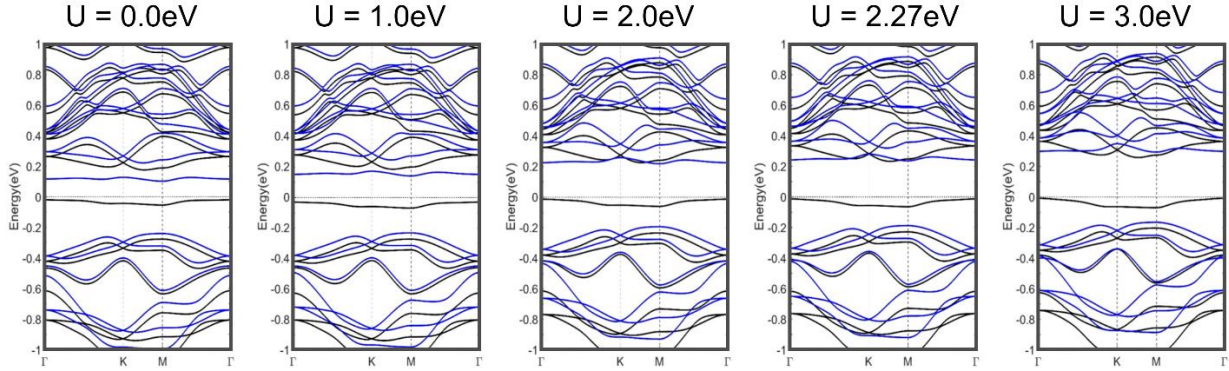

**Supplementary Figure 6: Evolution of the GGA+U band structure with increasing values of Hubbard U from 0 to 3 eV.**

In the 1T-TaS$_2$ monolayer, the DFT calculation shows an exchange gap at the Fermi level, even without including a Hubbard U. However, this gap is much smaller than that measured experimentally. To reproduce experimental observations, a Hubbard U is added to the calculation to encompass Coulomb repulsion between electrons in neighboring CDW stars. Increasing the value of U enhances the gap at the Fermi level. The value of U=2.27eV, previously determined from linear response theory, is used in the calculations given in the main text.

## CDW, Exchange, and Mott gaps

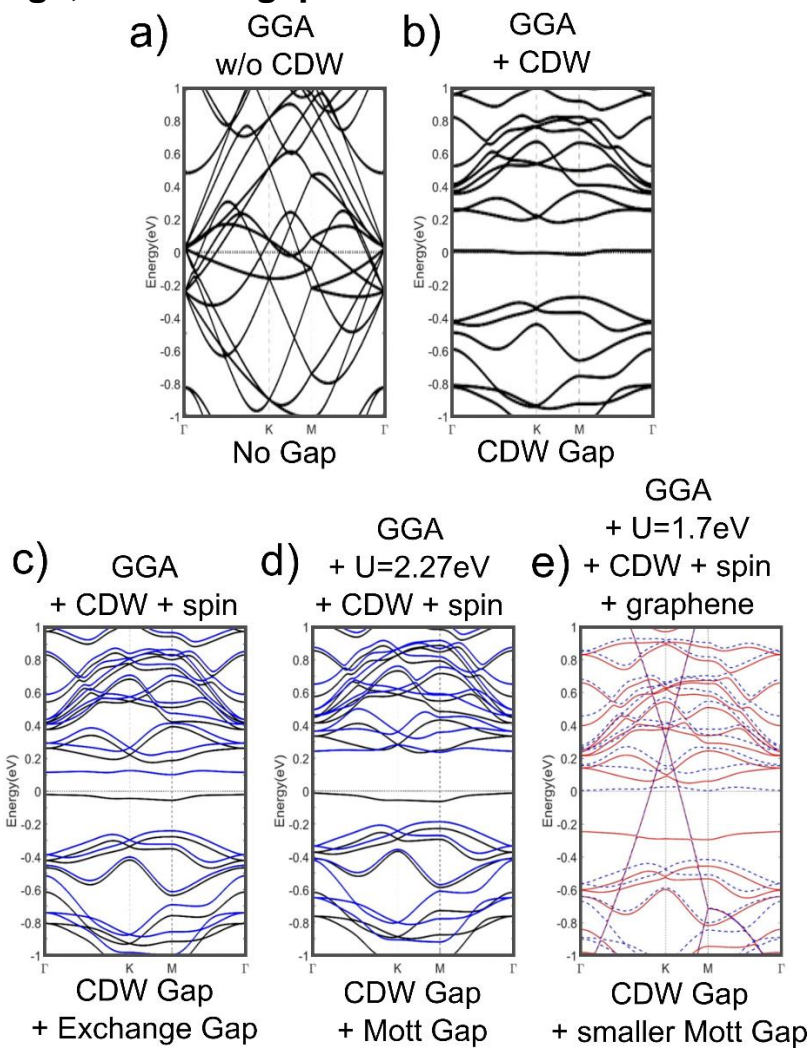

**Supplementary Figure 7:** (a) GGA band structure of undistorted 1T-TaS$_2$ shows no band gap. (b) After relaxing the lattice, a CDW gap opens with an in-gap, half-filled, isolated flat band stuck at E$_F$. (c) Adding spin splits the isolated flat band into two spin-polarized bands separated by an exchange gap. (d) Increasing the on-site Hubbard U expands the exchange gap into the experimentally observed Mott gap. (e) With the addition of graphene, the Hubbard U and the associated Mott gap are reduced in size due to screening from the graphene layer.

The outline of the emergence of the insulating state of 1T-TaS$_2$ begins with the CDW lattice reconstruction which isolates a narrow, flat band at the Fermi level within the CDW-induced gap. Including spin freedom and exchange interactions, the flat band splits into an exchange gap at the Fermi level. Adding a Hubbard U, enhances the gap into a Mott gap which reproduces experimental observations. Finally, adding a graphene layer on top causes a shift of the 1T-TaS$_2$ bands toward lower energies and adds a Dirac cone, centered at the K-points of the superstructure Brillouin zone. Additionally, the graphene layer is found to reduce the spacing between the Upper and Lower Hubbard bands of 1T-TaS$_2$ which is captured by using a smaller value of Hubbard U=1.70eV.

# BAND STRUCTURE OF GRAPHENE/1T-TAS$_2$

## Comparing DFT Unit Cells

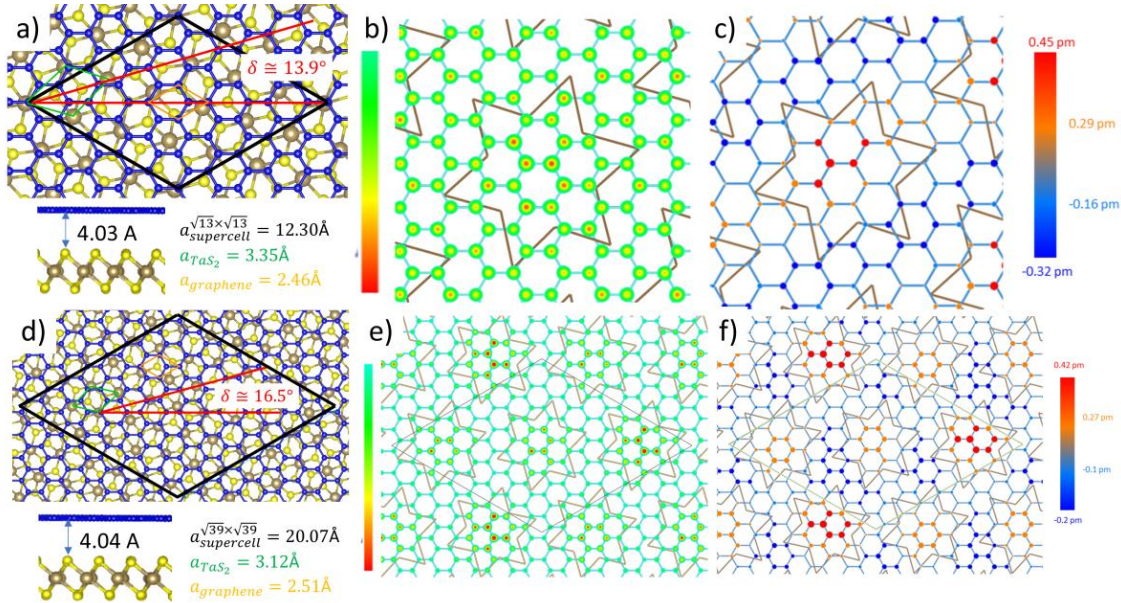

**Supplementary Figure 8: a)** Top view and side view of G/TaS$_2$ heterostructure with the 5x5 G/$\sqrt{13} \times \sqrt{13}$ TaS$_2$ R13.9$^0$ supercell containing 89 atoms with one star used in the DFT calculation (reproduced from main text Fig. 2d). **b)** DFT calculated charge transfer of the graphene layer based the 5x5 G/$\sqrt{13} \times \sqrt{13}$ TaS$_2$ R13.9$^0$ supercell shows a local modulation of doping with the periodicity of the CDW in TaS$_2$. Color scales range from -4 10$^{-4}$e/ Å$^3$ (red) to -2 10$^{-4}$e/ Å$^3$ (green). **c)** Out-of-plane displacement of Graphene induced by TaS$_2$ CDW based on the 5x5 G/$\sqrt{13} \times \sqrt{13}$ TaS$_2$ R13.9$^0$ supercell. **d)** Top view and side view of G/TaS$_2$ heterostructure with the 8x8 G/$\sqrt{39} \times \sqrt{39}$ TaS$_2$ R16.5$^0$ supercell containing 245 atoms with 3 stars used in the DFT calculation. **e)** DFT calculated charge transfer of the graphene layer based the 8x8 G/$\sqrt{39} \times \sqrt{39}$ TaS$_2$ R16.5$^0$ supercell shows a local modulation of doping with the periodicity of the CDW in TaS$_2$. Color scales range from -5 10$^{-4}$e/ Å$^3$ (red) to -3 10$^{-4}$e/ Å$^3$ (light blue). **f)** Out-of-plane displacement of Graphene induced by TaS$_2$ CDW based on the 8x8 G/$\sqrt{39} \times \sqrt{39}$ TaS$_2$ R16.5$^0$ supercell. The black, green, and orange diamonds in (a) and (d) represent the G/TaS$_2$ supercell, TaS$_2$ unit cell, and graphene unit cell, respectively. The lattice constants indicated in (a) and (d) are determined by DFT fully relaxation calculations. The interlayer distances between TaS$_2$ and Graphene are 4.03 Å (a) and 4.04 Å (d) in the minima 5x5 G/$\sqrt{13} \times \sqrt{13}$ TaS$_2$ R13.9$^0$ supercell and 8x8 G/$\sqrt{39} \times \sqrt{39}$ TaS$_2$ R16.5$^0$ supercell, respectively.

We have repeated the DFT calculations based on the minima 5x5 G/$\sqrt{13} \times \sqrt{13}$ TaS$_2$ R13.9$^0$ supercell containing 89 atoms with 1 CDW star as shown in the main text (reproduced here in **Supplementary Figure** 8a). As shown there, the largest induced local charge transfers and lattice displacements in graphene are located around the center of the TaS$_2$ CDW star (reproduced in S8b and S8c, respectively). Overall, the induced charge modulations and lattice distortions in graphene match well with the TaS$_2$ CDW structure as observed in our STM measurement. Furthermore, we perform the same calculations for an even larger 8x8 G/$\sqrt{39} \times \sqrt{39}$ TaS$_2$ R16.5$^0$ supercell containing 245 atoms with 3 TaS$_2$ CDW stars as shown in **Supplementary Figure** 8d. Fig. S8e illustrates the charge transfer in the graphene layer induced by TaS$_2$ showing a local modulation of doping with the periodicity the same as the CDW in TaS$_2$. The charge modulation obtained from this large supercell not only agrees with our experimental findings but also consists with Fig. 1(c) (reproduced in S8b) based on the minima supercell. Similar displacement behavior can be seen in Fig. S8f that graphene distorts upward around the center of the CDW stars while distorts downward otherwise. We note that the vertical displacement above 0.1pm is generally larger than the horizontal displacement below 0.1pm. We can draw several important conclusions from these results.

With these 2 calculations showing the same trend, we thus are confident that our DFT accuracy is good enough to give a consistent picture of the lattice distortion in graphene as induced by $TaS_2$ CDW. Additionally, as DFT calculations show the same charge modulation and lattice distortion trend in graphene for both the minima 5x5 G/$\sqrt{13} \times \sqrt{13}$ $TaS_2$ R13.9$^0$ supercell and the large 8x8 G/$\sqrt{39} \times \sqrt{39}$ $TaS_2$ R16.5$^0$ supercell, we are confident to conclude that the CDW-shape charge density distribution and lattice distortion in graphene is insensitive to the supercell size adopted in DFT calculations.

Further, we can conclude that the twist angle used in the calculation (13.9$^0$) is not a requirement for the CDW proximity effect observed in the DFT calculation. The fact that the CDW proximity effect is seen in both DFT calculations and STM measurements for multiple twist angles, leads us to believe that the CDW proximity effect is not twist-angle dependent.

## PROXIMITY INDUCED CDW GAP

To resolve the CDW gap induced in graphene by $TaS_2$, we performed high-accuracy band structure calculations for G/$TaS_2$ (Fig. S9(a)) and graphene with the same structure but with $TaS_2$ removed as shown in Fig. S9(b,c) and Fig. S9(d), respectively. As indicated by the green circle in Fig. S9(b), the Dirac point is shifted to ~0.3 eV above Ef because of the charge transfer from graphene to $TaS_2$. Due to the CDW proximity effect from $TaS_2$, the graphene Dirac point indeed opens up a CDW gap with the gap size of 1.7~1.8 meV as shown in Fig. S9(c) which is too small to observe at our experimental temperatures of 77K. For the band structure of graphene in the same lattice structure with $TaS_2$ layer removed presented in Fig. S9(d), the free standing graphene with the CDW distortion also gives a CDW gap of 1.6 meV at the Dirac point. Note that we did not include spin-orbit coupling (SOC) in these calculations. Therefore, we can conclude that this gap at the Dirac point is solely induced by the CDW lattice distortion in graphene via CDW proximity from $TaS_2$. On the other hand, we calculated band structure with SOC included for undistorted graphene, which gives a SOC gap in the order of 0.01 meV consistent with literature data. Thus we can rule out both the SOC effect and the numerical errors as the origin in the formation of the graphene CDW gap. The small size of the proximity CDW gap around 1.7 meV in graphene is much smaller than the CDW gap size around 0.4 eV in $TaS_2$. This is reasonable owing to the weak van der Waals interaction-mediated CDW proximity effect. At our experimental temperature of 77K this tiny gap would be smeared out by thermal fluctuations. Even if the experiment were performed at much lower temperatures such a tiny CDW gap in graphene would be very difficult to distinguish from the strong signal of the 1T-$TaT_2$ layer.

To examine the $TaS_2$ stacking effect on the graphene CDW gap, we further performed calculations for G/$TaS_2$ bilayer with AA-stacking and AL-stacking as shown in Fig. S9(e-h) and Fig. S9(i-l), respectively. Different stackings affect the induced graphene CDW gap slightly with the gap in meV order ranging from 0.9 meV to 1.9 meV compatible with each other. With the TaS2 layers removed, the distorted graphene exhibits CDW gap at the Dirac point of 1.6, 4.0, and 4.1 meV compatible with other as well as with the G/TaS2 cases. In all cases, the induced graphene CDW gap survives with different $TaS_2$ stackings.

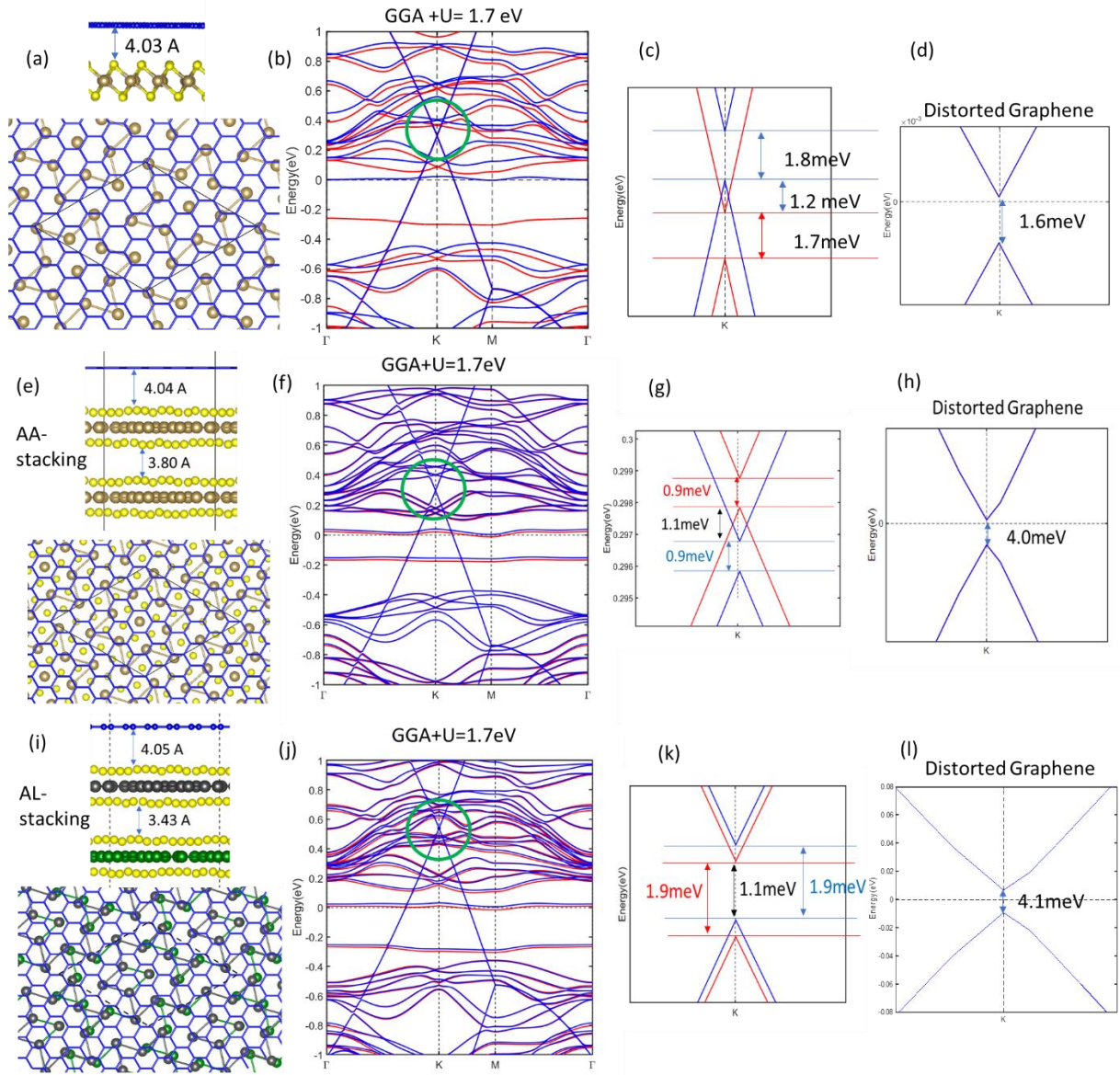

**Supplementary Figure 9 (a)** Side view and top view of G/TaS2 monolayer discussed in the main text. **(b)** Band structure of G/TaS2. Red and blue curves denote spin up and spin down bands, respectively. **(c)** Zoom-in plot of the Dirac bands of (b). The proximity CDW gap of ~1.7 meV is induced at the Dirac point. **(d)** Band structure of CDW-distorted graphene with TaS2 layer removed. A CDW gap of size 1.6 meV can also be seen at the graphene Dirac point. **(e-h)** Counterparts of (a-d), respectively, for G/TaS2 bilayer with AA-stacking. **(i-l)** Counterparts of (a-d), respectively, for G/TaS2 bilayer with AL-stacking. All the lattice structures are from DFT fully relaxation calculations. In all G/TaS2 heterostructures considered, the graphene CDW gaps at the Dirac point are in meV order ranging from 0.9 meV to 1.9 meV compatible with each other. With the TaS2 layers removed, the distorted graphene exhibits CDW gap at the Dirac point of 1.6, 4.0, and 4.1 meV compatible with other as well as with the G/TaS2 cases.

# ADDITIONAL BIAS DEPENDENT TOPOGRAPHY

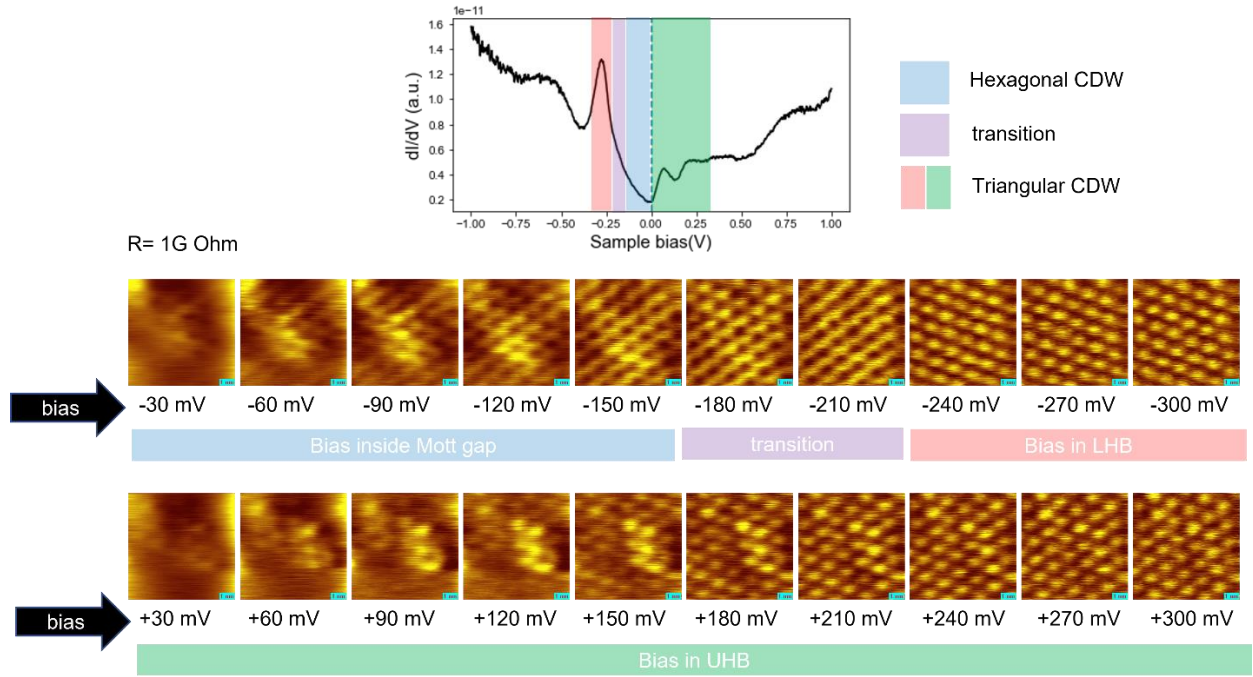

**Supplementary Figure 10** Bias dependent topography scans on Graphene/TaS2 sample in the bias range [ -300 mV, +300 mV]. Top row shows the measured dI/dV spectrum and the observed lattice pattern for various bias ranges. The middle row shows the evolution of the topography for negative biases while the bottom row shows the same for positive biases. The CDW pattern appears triangular for all biases outside the mott gap and hexagonal within the band gap. There is a narrow transition regime where it is difficult to differentiate the two. The topography signal is weak for |bias|<90 mV. All scans were recorded while keeping the junction resistance R= 1 giga Ohm.

Since we see the Hubbard bands in the Graphene/1T-TaS$_2$ spectra, it is safe to assume that the Mott gap in the 1T-TaS$_2$ survives when covered with Graphene. If we now set our bias set point inside this gap ( $V_b < (LHB_{center} + LHB_{width}) \approx -272 + 102 = -170\ mV$), then there are no states originating from the TaS$_2$ inside the bias window. As a result, the topography signal should only measure the states in the Graphene layer.

For any bias outside this narrow range, there are always TaS$_2$ states available to tunnel into. As a result, the topographic contribution from the Graphene cannot be isolated and substantial direct tunneling into the TaS$_2$ can coexist. For positive biases, the typical total number of states in the bias window for TaS$_2$ is 40-60 times larger than that in Graphene. Therefore, a large fraction of the total tunneling current goes directly into the TaS$_2$. For negative biases the situation is far better.

We therefore study how the topography changes as a function of bias set-point. These are the main observations-

1. For Vb<-210 mV and for Vb>0 mV we see a triangular CDW lattice. This is expected because the bias windows contain one of the Hubbard bands. We know from Partial LDOS calculations and from the experiment that the Hubbard bands have a higher weight near the centers of the stars-of-David compared to the sides resulting in a triangular pattern.

2. For -150 mV< Vb < 0 we observe a weak hexagonal lattice. A hexagonal lattice can be thought of as the inverse of a triangular lattice. From the argument above, we know that the topography in this bias range only consists of direct tunneling into the Graphene layer.

From the second-order perturbation model as well as from DFT, we showed that the CDW order parameter should be out of phase in the Graphene layer compared to the CDW order parameter in the TaS$_2$ layer. As a result, we can expect a lower density of states in the Graphene directly on top of the SD centers compared to the sides. This would correspond to a hexagonal lattice in topography.

## CHEMICAL POTENTIAL VARIATION/SCREENING MODEL

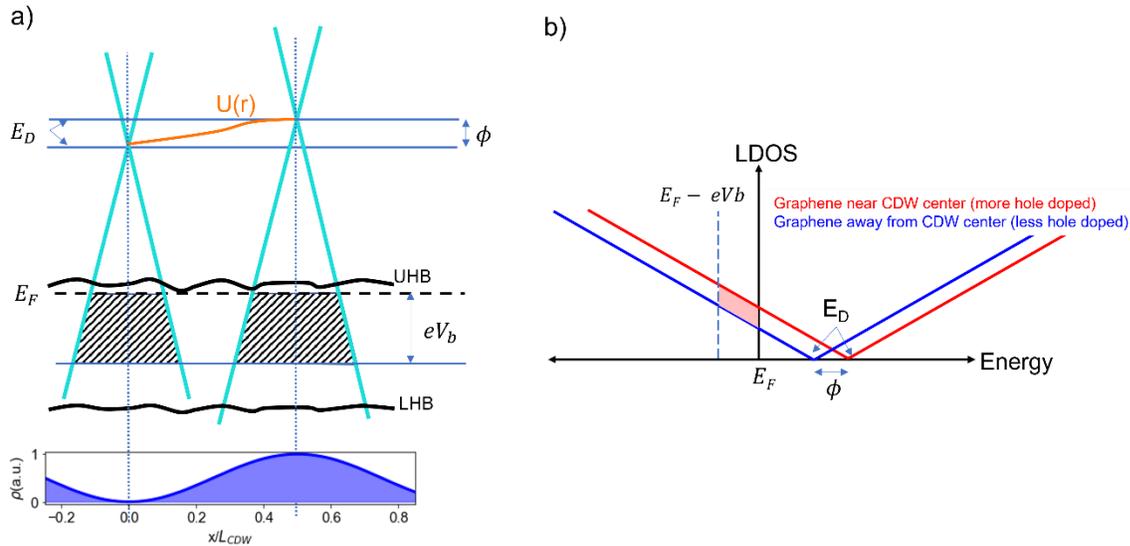

**Supplementary Figure 11: Expected effects of screening. (a)** A schematic of the Graphene Dirac cone located on top of the central Ta atom in a SD (right) and away from the central Ta atom (left). The local chemical potential is modulated by the underlying CDW by an amount $U(r)$ reaching a maximum amplitude $\phi$. The excess negative charge density near the central Ta atom locally shifts the Dirac point higher in energy The hatched areas show the occupied states in a small energy window $eV_b$, below the Fermi level ($E_F$) located near the UHB. There are more occupied states in the Graphene near the center of the SD than away from it. **(b)** The same is shown as a local density of states plot where the curves represent the Graphene states near the center of the SD (red) and far away from it (blue). The shaded area represents the excess occupied states in a small energy window $eV_b$, below the Fermi level ($E_F$).

The electrons in Graphene can respond to an underlying external potential and partially screen it. This locally changes the chemical potential of Graphene. The charge density wave in 1T-TaS$_2$ should act as a very weak periodic potential $U(r)$ for the Graphene layer in contact with it. Note that this potential is expected to fall off quickly with distance from the surface because the negative CDW charge is balanced by the positive charge of the TaS$_2$ lattice.

Let's assume that there is no hybridization between the Graphene and TaS$_2$ bands. The average doping of the Graphene layer (~0.3 eV) is set by the differences in the work function of Graphene and TaS$_2$. The CDW potential should locally modulate this average doping level. The Graphene close to the SD centers should have excess hole doping compared to the Graphene away from the SD center in response to the excess negative charge density there (**Supplementary Figure 11a**). This would lead to the Dirac point moving up in energy by an amount $\phi$. For a small bias window ($eV_b$) in the Mott gap of TaS$_2$, this leads to an excess of occupied states near the center of the SD (**Supplementary Figure** 11b). STM topography scans measured at this bias should show a CDW-like pattern in Graphene which peaks on top of the SD center. This should result in a triangular CDW pattern which is in-phase with the $TaS_2$ CDW. This does not match the observed STM topography. This means that local chemical potential variation/screening cannot explain the proximity induced CDW in Graphene. This points to the role played by band hybridization as the dominant mechanism.

# EFFECTS OF DIRECT TUNNELING INTO TAS$_2$ LAYER

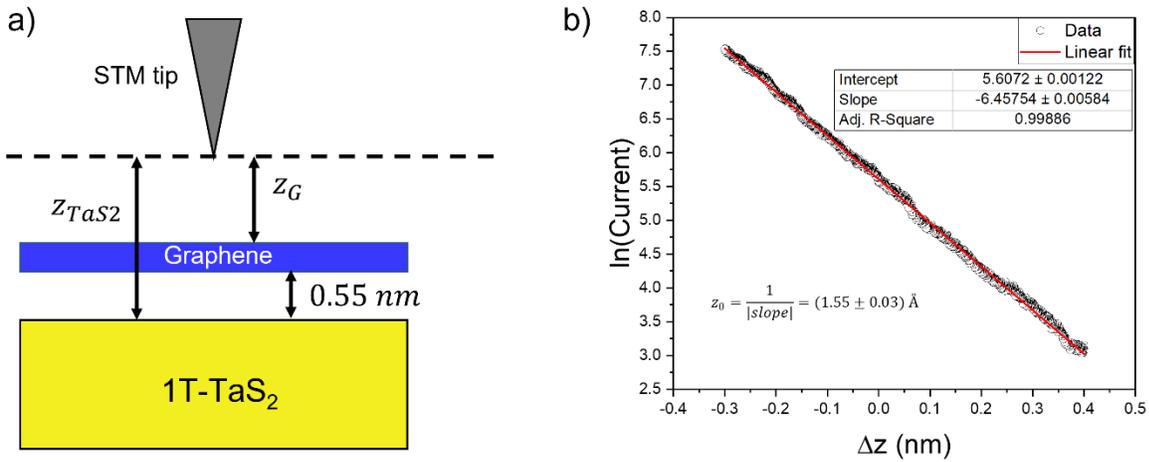

**Supplementary Figure 12: parallel tunneling into TaS$_2$ layer. (a) Cartoon showing the tunneling junction. The STM tip is at a distance $z_G$ away from the Graphene layer and $z_{TaS2}$ away from the TaS$_2$ layer. The two layers are separated by approximately 0.55 nm. Tunneling current can flow between the tip and both layers but is exponentially suppressed with increasing distance. (b) Extracting the tunneling current decay constant $z_0$ using IZ spectroscopy. The current vs z displacement is plotted on a log-linear plot. The slope of the line is extracted using a linear fit (red line).**

STM tunneling current is proportional to the number of available states in the bias window and decays exponentially with z distance $[I(z, V_b) \propto e^{-\frac{z}{z_0}} \int_0^{eV_b} d\epsilon \, DOS(\epsilon)]$. For Graphene on TaS$_2$, we can estimate the tunneling current which flows into the Graphene layer and TaS$_2$ layer using this simple picture as shown in **Supplementary Figure** 12a. The total tunneling current can be expressed as a sum of these currents as, $I \approx I_G + I_{TaS2}$ where,

$$I_G \propto e^{-\frac{z_G}{z_0}} \int_0^{eV_b} d\epsilon \, DOS_G(\epsilon)$$

and,

$$I_{TaS2} \propto e^{-\frac{z_{TaS2}}{z_0}} \int_0^{eV_b} d\epsilon \, DOS_{TaS2}(\epsilon)$$

Therefore,

$$\frac{dI}{dV_b} \propto e^{-\frac{z_{TaS2}}{z_0}} DOS_{TaS2}(\epsilon) + e^{-\frac{z_G}{z_0}} DOS_G(\epsilon)$$

$$\propto DOS_{TaS2}(\epsilon) + e^{\frac{z_{TaS2}-z_G}{z_0}} DOS_G(\epsilon)$$

Since the tip is closer to the Graphene compared to the TaS2 by $z_{TaS2} - z_G$ =0.55 nm, the current flowing into the Graphene is higher by a factor $\exp\left(\frac{z_{TaS2}-z_G}{z_0}\right) = \exp\left(\frac{5.56}{1.55}\right) \approx 39$. Here, $z_0 = 0.155$ nm is a function of the tunneling barrier height and was determined experimentally using IZ-spectroscopy (**Supplementary Figure** 12b).

On the other hand, the total density of states in the TaS$_2$ layer is typically higher than that in Graphene depending on the energy. For instance, according to the calculated DOS, for a bias voltage of -350 mV, there are about 3.19 states in the TaS$_2$ for every state in Graphene. As a result, the net tunneling current flowing into the Graphene layer is about 39/3.19=12.2 times that of TaS$_2$ for this bias. For large bias voltages of either sign, the DOS of TaS$_2$ is significantly larger than that of the Graphene layer. At these bias values, direct tunneling into the TaS$_2$ states cannot be ignored and significantly affects the STM topography scans.

On the other hand, the expected spectra are given by, $\frac{dI}{dV} \propto DOS_{TaS2} + 39 * DOS_{Graphene}$. Note that this scaled sum is significantly different from the naïve direct sum: $\frac{dI}{dV} \propto DOS_{TaS2} + DOS_{Graphene}$. We plot both the direct sum and scaled sum in Main figure 2d. The scaled sum qualitatively matches the experimentally observed spectrum (Main Figure 2f) much better than a direct sum.

## MEAN-FIELD THEORY APPROACH

We find that the proximity induced CDW is a novel effect driven by periodically varying band hybridization. This can be modeled using a simplified Hamiltonian of the graphene/TaS$_2$ system given by:

$$H = H_d + H_c + H_t,$$

$$H_d = \sum_{\langle i,j\rangle,\sigma} -t_{ij}^d d_{i,\sigma}^\dagger d_{j,\sigma} + h.c. - \sum_{\langle i',j'\rangle,\sigma} (\Delta_d^{CDW}(i',j'))^* d_{i',\sigma}^\dagger d_{j',\sigma} + h.c. + \sum_{\langle i',j'\rangle} |\Delta_d^{CDW}(i',j')|^2,$$

$$H_c = \sum_{\langle i,j\rangle,\sigma} -t_{ij}^c c_{i,\sigma}^\dagger c_{j,\sigma} + h.c. = \sum_{k,\sigma} (\epsilon_k - \mu) c_{k,\sigma}^\dagger c_{k,\sigma},$$

$$H_t = -t \sum_{i,\sigma} c_{i,\sigma}^\dagger d_{i,\sigma} + h.c.,$$

where $H_d$ ($H_c$) stands for the simplified Hamiltonian of the 1T-TaS$_2$ (graphene) layer, respectively, and $H_t$ describes a weak charge transfer (hopping) term between these two layers (we neglect a small mismatch in spatial locations between the nearest-neighbor sites of the corresponding layer). We assume that CDW order already exists on the insulating TaS$_2$ layer via electron-phonon coupling with the order parameter $\Delta_d^{CDW}(i',j') \equiv \sum_\sigma \langle d_{i',\sigma}^\dagger d_{j',\sigma}\rangle$ where $i', j'$ are sites within the CDW unit cell. The graphene layer has the linear Dirac dispersion: $(\epsilon_k - \mu) \approx \hbar v_F |k - k_F|$ where $v_F$, $k_F$ are the Fermi velocity and Fermi wave-vector respectively. Here, $i, j$ refer to the nearest-neighbor sites of the corresponding lattices, and $t_{ij}^{d(c)}$ refers to the nearest-neighbor tight-binding hoping terms on the 1T-TaS$_2$ (graphene) layer, respectively.

Via the second order perturbation in the $H_t$ term of the CDW unit cell, the following short-ranged exchange term $H_t^{(2)}$ between electrons on the two layers is generated:

$$H_t^{(2)} = t^2 \sum_{\langle i',j'\rangle,\sigma,\sigma'} c_{i',\sigma}^\dagger d_{i',\sigma} d_{j',\sigma'}^\dagger c_{j',\sigma'} + h.c.$$

A simple mean-field decoupling of $H_t^{(2)}$ in terms of $\Delta_d^{CDW}(i',j')$ (considering only $\sigma = \sigma'$ and assuming spin-isotropic CDW order $\langle d_{j',\uparrow}^\dagger d_{j',\uparrow}\rangle = \langle d_{j',\downarrow}^\dagger d_{j',\downarrow}\rangle$) gives $H_t^{(2)} \to H_{t^2}^{MF}$ with:

$$H_{t^2}^{MF} \approx -t^2/2 \sum_{\langle i',j'\rangle,\sigma} (\Delta_d^{CDW}(i',j'))^* c_{i',\sigma}^\dagger c_{j',\sigma} - t^2/2 \sum_{\langle i',j'\rangle,\sigma} (\Delta_d^{CDW}(i',j'))^* \langle c_{i',\sigma}^\dagger c_{j',\sigma}\rangle + h.c.,$$

where the mean-field decoupling term $\langle c_{i',\sigma}^\dagger c_{j',\sigma}\rangle d_{i',\sigma}^\dagger d_{j',\sigma}$ in $H_t^{(2)}$ is neglected since we expect $\left|\langle c_{i',\sigma}^\dagger c_{j',\sigma}\rangle\right| \ll \left|\langle d_{i',\sigma}^\dagger d_{j',\sigma}\rangle\right|$. The CDW proximity effect is manifested in $H_{t^2}^{MF}$ as a weak CDW order $\sum_\sigma \langle c_{i',\sigma}^\dagger c_{j',\sigma}\rangle$ is induced on the graphene layer by the weak second-order charge transfer between the two layers with the following identification:

$$\Delta_c^{CDW}(i',j') \equiv -1/2 \sum_\sigma \langle c_{i',\sigma}^\dagger c_{j',\sigma}\rangle^* = -t^2/2 \left(\Delta_d^{CDW}(i',j')\right)^*,$$

or equivalently, $\sum_\sigma \langle c_{i',\sigma}^\dagger c_{j',\sigma}\rangle = t^2 \Delta_d^{CDW}(i',j')$. Via the above identification, the Hamiltonian $H_{t^2}^{MF}$ can be expressed as:

$$H_{t^2}^{MF} = \sum_{\langle i',j'\rangle,\sigma} \Delta_c^{CDW}(i',j') c_{i',\sigma}^\dagger c_{j',\sigma} + h.c. + 2|\Delta_c^{CDW}(i',j')|^2,$$

which leads to $\Delta_c^{CDW}(i',j') = -1/2 \sum_\sigma \langle c_{i',\sigma}^\dagger c_{j',\sigma}\rangle^*$ identified above via minimizing the free energy associated with $H_{t^2}^{MF}$ with respect to $\Delta_c^{CDW}(i',j')$. Note that from above derivations, we indeed find that $\left|\langle c_{i',\sigma}^\dagger c_{j',\sigma}\rangle\right| \sim t^2 \left|\langle d_{i',\sigma}^\dagger d_{j',\sigma}\rangle\right| \ll \left|\langle d_{i',\sigma}^\dagger d_{j',\sigma}\rangle\right|$, as expected. Note also that the CDW order parameters induced on graphene layer shows the opposite sign with respect to that on 1T-TaS$_2$ layer, consistent with the hole-like (particle-like) CDW intensity on graphene (1T-TaS$_2$) layer obtained from DFT calculations, respectively. Meanwhile, the opposite sign of the proximity induced CDW on the graphene layer with respect to the CDW on the TaS$_2$ layer as predicted within our mean-field theory is in agreement with the "out-of-phase" relation between CDW patterns observed inside and outside of Mott gap by our STM measurement (see Fig. 2 g-i and corresponding text above). In addition, the mean-field theory predicts that the amplitude of the CDW in graphene is much weaker than that in 1T-TaS$_2$, again consistent with the DFT and the STM results. These findings provide strong support for the CDW proximity effect in graphene. We emphasize here that the above mechanism based on charge transfer is distinct from all the previously realized proximity effects, including superconducting, magnetic, and spin-orbit proximity effects.